\documentstyle[12pt]{article}

\begin{document}
\begin{titlepage}
\title{Motion of Curves on Two Dimensional Surfaces and Soliton Equations}
\vspace{5cm}
\author{Metin G{\" u}rses \\
{\small Department of Mathematics, Faculty of Science} \\
{\small Bilkent University, 06533 Ankara - Turkey}\\
email:gurses@fen.bilkent.edu.tr}
\maketitle
\begin{abstract}
A connection is established between the soliton
equations and curves moving in a three dimensional space $V_{3}$.
The sign of the self-interacting terms of the soliton equations
are related to the signature of $V_{3}$. It is shown that there corresponds
a moving curve to each soliton equations.

\end{abstract}
\end{titlepage}

Differential geometry and partial differential equations (PDEs) are two
different  research areas in mathematics . When we study
some local properties of surfaces in Euclidean ($E_{3}$) or Minkowskian
($M_{3}$) 3-spaces
we face with some known PDEs. For instance the Liuoville and Sine-Gordon
equations describe surfaces of constant Gaussian curvature \cite{EIS}.
Gauss-Codazzi-
Mainardi equations describe the surfaces  embedded in $E_{3}$ or in $M_{3}$.
These equations are used for the construction of
the soliton connection \cite{CRA},\cite{GUR1},\cite{KON}.
Here the differential geometrical tools are utilized to find
for example the Backlund transformations and prolongation structures
\cite{EST} of the soliton equations.

During the last two decades another virtue
of the differential geometry arised in soliton theory. The Serret-Frenet
equations for the family of curves (the motion of curves) give
certain coupled partial differential equations for the curvature ($k$)
and torsion ($\tau$)
scalars of these curves \cite{HOS}-\cite{per}. It was shown
that some soliton equations like the modified Korteweg
de Vries (mKdV), sine-Gordon and nonlinear Schrodinger
(NLS) are among the equations which may arise from the motion
of space curves.
All these considerations were in Euclidean three space $E_{3}$.
This is why only one
versions of the nonlinear couplings of the mKdV and NLSEs could have been
obtained.

In this work we take a three space $V_{3}$ with signature
$1+2\epsilon$ , where $\epsilon^2=1$. This means that curves
in $M_{3}$ will also be considered. Self interacting terms in the
evolution equations of the curvature and the torsion of these
curves depend upon signature of the space $V_{3}$. The sign
difference of the self interaction terms is due to
signature change of the three space. If for instance a curve $C$
is moving in $E_{3}$ (or in $M_{3}$) focusing (or defocusing) versions
of mKdV or NLS equations arise.

The motion of the curve $C$ is described by three functions $p$ , $q$ and
$w$. The function $w$ is determined in terms of the others
but the functions $p$ and $q$
are left arbitrary. Each choice of these functions gives a different
classess of curves in $V_{3}$. It is in principle possible to convert
the differential
equations satisfied by the scalars $k$ and $\tau$ to any system of coupled
two nonlinear PDEs. Here we should remark that not all these equations are
integrable. The integrability prorperty of these equations (for each
choice of $p$ and $q$) should be examined. The functions $p$ and $q$ can
be suitably choosen to make the evolution equations satisfied by $k$ and
$\tau$ integrable. So far , for this purpose  \cite{HOS}-\cite{per}
$p$ and $q$ were assumed to be local functions of $k$ and $\tau$.
By this way mKdV , NLS, and complex mKdV equations
could be obtained.

On the other hand one may obtain , by a proper choice of $p$ and $q$  (since
they free), all possible integrable equations. This can be done
by relaxing the locality assumptions on the functions $p$ and $q$.
Sine-Gordon equation is obtained by assuming that $q=\tau=0$ and
$p$ is a nonlocal function of the curvature $k$ \cite{khm}-\cite{lam}.
We show that any integrable system of coupled two nonlinear PDEs can
be obtained by assuming nonlocal functional dependence. In this way it
is possible to obtain
for instance the AKNS \cite{ABL} hierarchy. Hence in general there exist
a curve $C$ moving in a $V_{3}$ corresponding to any integrable nonlinear
differential equation (one or two coupled equations).

Some nonlinear partial differential equations , such as the sine-Gordon
and the Liuoville equations arise from the surfaces of constant
Gaussian curvature. Here we show that such equations and many others
may also arise from two dimensional surfaces with vanishing
Gaussian curvature , flat surfaces (see also \cite{sym}).

Let $V_{3}$ define a 3-dimensional flat space with the line element

\begin{equation}
ds^2=\eta_{\mu\, \nu}\,dx^{\mu}\, dx^{\nu}  \label{mnk1}
\end{equation}

\noindent
where $\mu,\nu=1,2,3$ , $x^{\mu}=(t,y,z)$ and $\eta_{\mu\, \nu}=
diag (1,\epsilon, \epsilon)$. If $\epsilon=1$ then $V_{3}=E_{3}$ is an
Euclidean three space and if $\epsilon=-1$ then $V_{3}=M_{3}$ is a
pseudo-Euclidean (Minkowskian) three space. Hence (\ref{mnk1}) explicitly
takes the form

\begin{equation}
ds^2=dt^2+\epsilon\,dy^2+\epsilon\,dz^2  \label{mnk2}
\end{equation}

\noindent
Let $S$ be a surface in $V_{3}$ parametrized by $x^{\mu}(u,v)$,
and  let $C$ be a curve on $S$ defined by ${\bf \alpha}:I \rightarrow S$
and  parametrized by
its arclength $s \in I$. An orthonormal frame $(t^{\mu},n^{\mu},b^{\mu})$ at each
point of $C$ is defined by (recalling that  $x^{\mu}_{,s}= t^{\mu}$)

\begin{equation}
\eta_{\mu\, \nu}\,t^{\mu}\, t^{\nu}=1~,~
\eta_{\mu\, \nu}\,n^{\mu}\, n^{\nu}=\epsilon~,~
\eta_{\mu\, \nu}\,b^{\mu}\, b^{\nu}=\epsilon
\end{equation}

\noindent
all the other products vanish.
The Serret-Frenet equations are ($x^{\mu}_{,s}= t^{\mu}$)

\begin{eqnarray}
t^{\mu}_{,s}=k\, n^{\mu}, \label{sf1}\\
n^{\mu}_{,s}=-\epsilon\,k\,t^{\mu}-\tau\,b^{\mu}, \label{sf2}\\
b^{\mu}_{,s}=\tau\, n^{\mu} \label{sf3}
\end{eqnarray}

\noindent
where $k$ and $\tau$ are the curvature and the torsion scalars of the
curve $C$ at any point $s$. The vectors
$t^{\mu}$, $n^{\mu}$ and $n^{\mu}$ are respectively the tangent ,
normal and bi-normal vectors to the curve at any point $s$ \cite{CAR}

A curve on $S$ is given by $\alpha^{\mu}(s)=x^{\mu}(u(s),v(s))$.
This curve may be considered as a member of the family of curves
$\beta^{\mu}_{\sigma}=x^{\mu}(u(s,\sigma),v(s,\sigma))$ for a fixed value
of $\sigma$. The change (motion) of the curve with respect to the parameter
$\sigma$ (on $S$) is given by

\begin{equation}
x^{\mu}_{,\sigma}=p\,n^{\mu}+w\,t^{\mu}+q\, b^{\mu} \label{mot1}
\end{equation}

\noindent
where the $p$ , $q$ and $w$ are functions of $s$ and $\sigma$. By using
the equation $x^{\mu}_{,s}=t^{\mu}$ and (\ref{mot1}) we get
$w_{,s}=\epsilon\,k\,p$ and $t^{\mu}_{,\sigma}$ (partial derivative of
the vector
$t^{\mu}$ with respect to $\sigma$). Using $t^{\mu}_{,\sigma}$ obtained this
way and the first equation (\ref{sf1}) of the Serret-Frenet equations one
obtains $k_{,\sigma}$ and $n^{\mu}_{,\sigma}$. Following the similar
approach one finds derivatives of the scalars ($k$,$\tau$) and vectors
($t^{\mu},n^{\mu}$, $b^{\mu}$). They are given by

\begin{equation}
t^{\mu}_{,\sigma}=(p_{,s}+kw+\tau\,q)\,n^{\mu}+
(q_{,s}-\tau\,p)\,b^{\mu} \label{mt1}
\end{equation}

\begin{eqnarray}
n^{\mu}_{,\sigma}=-\epsilon\,(p_{s}+kw+\tau\,q)\,t^{\mu}+
{1 \over k}\,[(q_{,s}-\tau\,p)_{,s}- \nonumber \\
\tau\,(p_{,s}+kw+\tau\,q)]\,b^{\mu} \label{mt2}
\end{eqnarray}

\begin{eqnarray}
b^{\mu}_{,\sigma}=-{1 \over k}\,[(q_{,s}-\tau\,p)_{,s}-\tau\,(p_{,s}+kw+
\tau\,q)]\,n^{\mu}- \nonumber \\
\epsilon\,(q_{,s}-\tau\,p)\,t^{\mu} \label{mt3}
\end{eqnarray}

\noindent
The compatibility conditions give $w_{,s}=\epsilon\,k\,p$ and

\begin{equation}
k_{,\sigma}=(p_{,s}+kw+\tau\,q)_{,s}+
\tau\,(q_{s}-\tau\,p) \label{eq1}
\end{equation}

\begin{equation}
\tau_{,\sigma}=-[{1 \over k}\,[(q_{,s}-\tau\,p)_{,s}- \nonumber \\
\tau\,(p_{,s}+kw+\tau\,q)]]_{,s}
-\epsilon\,k(q_{,s}-\tau\,p) \label{eq2}
\end{equation}

\noindent
These equations may be written in a compact form

\begin{equation}
\left(\begin{array}{c}
k\\ \tau
\end{array} \right)_{,\sigma}
=\cal{R} \left(\begin{array}{c}
p\\ q
\end{array} \right) \label{ev1}
\end{equation}

\noindent
where

\begin{equation}
\cal{R}=\left(\begin{array}{clcr}
D^2+\epsilon\,k^2-\tau^2+\epsilon\,k_{,s}\,D^{-1}\,k &
(D\,\tau+\tau\,D)\\
D[{1 \over k}\,( D\tau+\tau\,D)+\epsilon\, \tau D^{-1}k]
+\epsilon\,k\tau & -D[{1 \over k}\, D^2-{\tau^2 \over k}]-\epsilon\,kD
\end{array} \right)  \label{sys1}
\end{equation}

\noindent
In the special case $\tau=q=0$ which means $C$ is a plane curve we have

\begin{equation}
k_{,\sigma}={\bf R}\,p \label{eq3}
\end{equation}

\noindent
where ${\bf R}$ is the recursion operator of the mKdV equation
$k_{,\sigma}={\bf R}\,k_{,s}$ given by

\begin{equation}
{\bf R}= D^{2}+\epsilon\,k^{2}+\epsilon\,k_{,s}\,D^{-1}\,k \label{eq4}
\end{equation}

\noindent
Here $D$ denotes the total derivative with respect to $s$ and $D^{-1}$ is
its inverse. Choosing , for instance $p=k_{,s}$ then Eq.(\ref{eq3}) reduces to
mKdV. The choices of the geometry $\epsilon=\pm 1$ we have focusing
and defocusing versions of the mKdV equations. Choosing
$p={\bf R}^{n}\,k_{,s}$ with $n=0,1,2,..$ we obtain the infinite
integrable hierarchy of the mKdV equations. As another local choices
we need to write
Eqs.(\ref{eq1}) and (\ref{eq2}) in a complexified from

\begin{eqnarray}
\phi_{,\sigma}=\{D^2+i\eta \, \epsilon\, \phi\,D^{-1}\, \tau \, \phi^{*}+
|\phi|^{2}+\phi_{,s}\,D^{-1}\, \phi^{*}\}\,(p\, \rho)+ \nonumber \\
\{-i \eta \, D^2-i\eta \, \epsilon\,|\phi|^{2}-
\epsilon\, \phi\, D^{-1}\, \tau\, \phi^{*}+
i \eta\,\epsilon\phi\,D^{-1}\, \phi^{*}_{,s}\}\,(q\, \rho)
\end{eqnarray}

\noindent
where $\eta^2=1$, $\rho=e^{i\eta\,(D^{-1}\, \tau)}$ and $\phi=k\, \rho$ and
$\phi^{*}$ is the complex conjugate of $\phi$. When $p=0$ and
$q=k$ then we have the nonlinear Schrodinger (NLS)
equation of both versions ($\epsilon=\pm 1$).

\begin{equation}
i\eta\, \phi_{,\sigma}=D^{2}\,\phi+{\epsilon \over 2}\, |\phi|^{2}\,
\phi
\end{equation}

\noindent
Another example is obtained by letting $p=k_{,s}$ and $q=-k \tau$. This
is the complex mKdV

\begin{equation}
\phi_{,\sigma}=D^{3}\,\phi+{3 \over 2}\, |\phi|^{2}\,\phi_{,s}
\end{equation}

\noindent
In all these choices the function
$p$ is choosen as local functions of the $k$. This means that $p$ is a
function of $k$ and its partial derivatives with respect to $s$ and
$\sigma$ to all orders. Other local choices of the function $p$ in terms
of $k$ may or
may not give integrable nonlinear partial differential equations (equations
admiting infinitely many generalised symmetries).
For each choice of $p$ one must check whether the resulting equation is
integrable \cite{FOK}, \cite{MIK}, \cite{GUR2}. The main motivation why
integrable equations are trying to be
choosen is their position in mathematics and physics.

It is also possible to choose the function $p$ as a nonlocal function of
$k$. Choosing for instantce $p={\bf R}^{-2}\,k_{,s}$ and letting
$k=\theta_{,s}$ we obtain the sine-Gordon
equation $\theta_{,s \,\sigma}=sin(\theta)$ \cite{lam}. Another choice
for instance may be $p={\bf R}^{-1}\, (R_{kdv})^{n}\,k_{,s}$ , where $R_{kdv}=
D^2+4\,k+2\,k_{,s}\,D^{-1}$ is the recursion operator of the KdV equation
$k_{,\sigma}=k_{,sss}+6\,k\,k_{,s}$. This choice will give the hierarchy of
the KdV equation $k_{,\sigma}=(R_{kdv})^{n}\,k_{s}$ ,
for $n=0,1,2,...$.
It is clear from these examples that
since $p$ is an arbitrary function,  Eq.(\ref{eq3}) may be reduced
to any nonlinear partial differential equation.
One can properly choose $p$ so that
all scalar integrable nonlinear PDE can be obtained from Eq.(\ref{eq3}).

In the general case by choosing $p$ and $q$ properly so that Eq.(\ref{sys1})
can be reduced to any system of coupled two nonlinear PDEs. As an example
letting

\begin{equation}
\left(\begin{array}{c}
p\\ q
\end{array} \right)
={\cal{R}}^{-1}\,(R_{akns})^{n} \left(\begin{array}{c}
k\\ \tau
\end{array} \right)_{,s} \label{ev2}
\end{equation}

\noindent
where $R_{akns}$ is the recursion operator of the AKNS system
of equations given by

\begin{equation}
R_{akns}=\left(\begin{array}{clcr}
D+2k\,D^{-1}\,\tau &
2k\,D^{-1} k\\
-2\tau\,D^{-1}\tau & -D-2\tau\,D^{-1}k
\end{array} \right)  \label{sys2}
\end{equation}

\noindent
Eq. (\ref{ev1}) reduces to AKNS hierarchy for $n=0,1,2,...$.Hence
there corresponds a class of moving curves in $V_{3}$ to each
system of two coupled soliton equations.

\begin{equation}
\left(\begin{array}{c}
k\\ \tau
\end{array} \right)_{,\sigma}
=(R_{akns})^{n} \left(\begin{array}{c}
k\\ \tau
\end{array} \right)_{,s} \label{ev3}
\end{equation}

\noindent
The derivatives of the vectors in the frame $e^{\mu}_{a}=
(t^{\mu},n^{\mu},b^{\mu})$
may be written in more familiar form
$d\, e^{\mu}_{a}=\Omega^{b}_{a}\,e^{\mu}_{b}$,
where in matrix notation $\Omega$ is a matrix valued 1-form. Here
$a,b=1,2,3$ and
$\Omega=\Omega_{s}\, ds+\Omega_{\sigma}\,d\sigma$,
where

\begin{equation}
\Omega_{s}=\left(\begin{array}{clcr}
0 & k & 0\\
-\epsilon\,k & 0 & -\tau\\
0 & \tau & 0
\end{array} \right)
~~,~~
\Omega_{\sigma}=\left(\begin{array}{clcr}
0 & w_{1} & w_{0}\\
-\epsilon\,w_{1} & 0 & w_{2} \\
-\epsilon\,w_{0} & -w_{2} & 0
\end{array} \right)
\end{equation}

\noindent
with

\begin{eqnarray}
w_{0}=q_{,s}-\tau\, p~~,~~
w_{1}=p_{,s}+kw+\tau \, q\\
w_{2}={1 \over k}\,[(q_{,s}-\tau\,p)_{,s}-
\tau\,(p_{,s}+kw+\tau\,q)]
\end{eqnarray}

\noindent
The 1-form $\Omega$ defines a connection with zero curvature.
This is due to the flatness of the space $V_{3}$.
Vanishing of the curvature of $\Omega$, i.e,
$d\Omega-\Omega \, \Omega=0$, is due to the evolution equations
given in (\ref{ev1}).
In order to compare this connection 1-form with the soliton connection
1-form we write it in more suitable form \cite{CRA},\cite{GUR1}

\begin{equation}
\Omega=\left(\begin{array}{clcr}
0 & \pi_{0} & \pi_{1}\\
-\epsilon\,\pi_{0} & 0 & \pi_{2} \\
-\epsilon\, \pi_{1} & -\pi_{2} & 0
\end{array} \right)
\end{equation}

\noindent
where the 1-forms $\pi_{0}$ , $\pi_{1}$ and $\pi_{2}$ are given by

\begin{eqnarray}
\pi_{0}=k\,ds+w_{1}\,d\sigma~~,~~
\pi_{1}=w_{0}\,d\sigma~~,~~
\pi_{2}=-\tau\,ds+w_{2}\,d\sigma
\end{eqnarray}

\noindent
These 1-forms satisfy (from the zero curvature condition)

\begin{eqnarray}
d\pi_{0}+\pi_{1}\, \pi_{2}=0~~,~~
d\pi_{1}- \pi_{0}\, \pi_{2}=0~~,~~
d\pi_{2}+\epsilon \pi_{0}\, \pi_{1}=0
\end{eqnarray}

\noindent
An $SL(2,R)$ valued soliton connection 1-form $\Gamma$ may be given
interms of the 1-forms $\pi_{0}$ , $\pi_{1}$ and $\pi_{2}$

\begin{equation}
\Gamma=\left(\begin{array}{clcr}
\theta_{0} & \theta_{1} \\
\theta_{2} & -\theta_{0}
\end{array} \right)
\end{equation}

\noindent
where

\begin{eqnarray}
\theta_{0}=\alpha\, \pi_{0}~~,~~
\theta_{1}=\alpha_{1}\,(\pi_{1}+{1 \over 2\alpha}\, \pi_{2})~~,~~
\theta_{2}=\alpha_{2}\,(\pi_{1}-{1 \over 2\alpha}\, \pi_{2}).
\end{eqnarray}

\noindent
Here we have $4\, \alpha^2+\epsilon=0$ , $\alpha_{1}\, \alpha_{2}=
\alpha^2\,$. Let $\Psi$ be a $2 \times 2$ matrix vaued (0-form)
function of $s$ and $\tau$. Then $d\Psi=\Gamma\, \Psi$ defines the Lax
equation without
a spectral parameter. Such a constant may be introduced by performing a
gauge transformation $\Gamma^{\prime}=S\, \Gamma\, S^{-1}+dS\, S^{-1}$.
Here $S$ is $2\times 2$ matrix valued function of $s$, $\sigma$ and the
spectral parameter. In this way we set up a correspondance between a curve $C$
moving in a space $V_{3}$ with a soliton connection.

\noindent
The line element (\ref{mnk1}) on the surface $S$, using the parameters
($s, \sigma$) of the moving $C$, reduces to

\begin{equation}
ds^2=(ds+w\, d\sigma)^2+\epsilon\,(p^2+q^2)\,d\sigma^2
\label{fm1}
\end{equation}

\noindent
The Gaussian
curvature $K$ of $S$ with the first fundamental form given in (\ref{fm1})
is different from zero in general. On the other hand by the choice
$\tau=q=0$ , the line element becomes

\begin{equation}
ds^2=(ds+w\,dt)^2+\epsilon\,p^2\, dt^2  \label{fm2}
\end{equation}

\noindent
The Gaussian curvature $K$ becomes

\begin{equation}
K={1 \over 4p}(k_{\sigma}-{\bf R}\,p)
\end{equation}

\noindent
which vanishes by virtue of the equation (\ref{eq3}). Hence all
the curves related to the eqn(\ref{eq3}) trace flat 2-surfaces.
It was usually believed that integrable equations arise from the curved
surfaces. For instance the sine-Gordon equation arise from the surface
with the line element
$ds^2=cos^2(\theta)\, d\sigma^2+sin^2(\theta)\, ds^2$,
which describe surfaces of constant negative Gaussian curvature \cite{EIS}.
Here we show that all integrable equations including the sine-Gordon equation
may also arise from flat 2-surfaces (for mKdV see \cite{sym}).

In this work we considered the motion of a curve in a three space $V_{3}$.
This condition
may be relaxed , but for an arbitrary $V_{n}$ where $n > 3$ the
evolution equations corresponding to the geometrical scalars
($k$, $\tau$,...)  of the curves
become quite complicated. It is perhaps more physical and significant
to consider the case $n=4$. This corresponds to classical
strings moving in four dimensional Minkowskian space. Hence
it will be quite interesting to see the correspondance between
strings and the soliton equations with four dependent variables.
Let $x^{\mu}(s,\sigma)$ denote the strings in $M_{4}$. In a
similar manner we define curves $x^{\mu}(s)$ parametrised with arclength
$s$ and its variations $x^{\mu}(s,\sigma)$. We have  the orthonormal tetrad
($t^{\mu},n^{\mu}, b_{1}^{\mu},b_{2}^{\mu}$) with

\begin{eqnarray}
\eta_{\mu \nu}t^{\mu}\, t^{\nu}=1~,~\eta_{\mu \nu}n^{\mu}\, n^{\nu}=\epsilon\\
\eta_{\mu \nu}b_{1}^{\mu}\, b_{1}^{\nu}=\epsilon~,~
\eta_{\mu \nu}b_{2}^{\mu}\, b_{2}^{\nu}=\epsilon
\end{eqnarray}

\noindent
where $\eta_{\mu \nu}=diag (1, \epsilon, \epsilon, \epsilon)$ , the
Greek letters run from 1 to 4.  Here $\epsilon = -1$ but we keep it
to compare the results obtained here with previous sections. The Serret-Frenet
equations governing the motion of the tetrad are as follows

\begin{eqnarray}
t^{\mu}_{,s}=k\, n^{\mu}, \label{sf4}\\
n^{\mu}_{,s}=-\epsilon\,k\,t^{\mu}-\tau_{1}\,b_{1}^{\mu}-
\tau_{2}\,b_{2}^{\mu}, \label{sf5}\\
b_{1,s}^{\mu}=\tau_{2}\, n^{\mu}+\tau_{3}\,b_{2}^{\mu} \label{sf6},\\
b_{2,s}^{\mu}=\tau_{1}\, n^{\mu}-\tau_{3}\,b_{1}^{\mu} \label{sf7}
\end{eqnarray}

\noindent
where $k, \tau_{1}, \tau_{2}, \tau_{3}$ are the geomerical scalars describing
the curvature , and torsions in each three space directions respectively.
Here we have $t^{\mu}=x^{\mu}_{,s}$ . Letting

\begin{equation}
x^{\mu}_{,\sigma}=p\,n^{\mu}+w\,t^{\mu}+q_{1}\, b_{1}^{\mu}+
q_{2}\,b_{2}^{\mu} \label{as1}
\end{equation}

\noindent
where $p,w, q_{1}, q_{2}$ are functions of $s$ and $\sigma$. Here it is
clear that when $\tau_{2},\tau_{3}$ and $q_{2}$ vanish we get the same
equations
obtained in the previous sections for three dimesional spaces. Hence
the strings in four dimensions has a very direct correspondance with the
integrable evolution equations. This connection and further progress on
the motion of curves in a four dimensional space will be communicated
elsewhere.

In this work we established a connection between the curves moving
in a three space with arbitrary signature (-1 or 3) and soliton equations.
We showed that to each soliton (integrable) equation there exists a class of
curve moving either in an Euclidean ($E_{3}$) or pseudo-Euclidean ($M_{3}$)
three spaces. The signature of $V_{3}$ and the sign of the self interacting
terms in the soliton equations are directly related.
We also showed that many integrable nonlinear PDEs may also
arise from flat surfaces contrary to the common belief so far \cite{sym}.

This work is partially supported by the Scientific and Technical Research
Council of Turkey and Turkish Academy of Sciences.

\end{document}